\documentclass[a4paper,english,aps,manuscript]{revtex4}
\usepackage[T1]{fontenc}
\usepackage[latin9]{inputenc}
\usepackage{textcomp}
\usepackage{graphicx}
\usepackage{esint}

\makeatletter

\pdfpageheight\paperheight
\pdfpagewidth\paperwidth

\@ifundefined{textcolor}{}
{%
 \definecolor{BLACK}{gray}{0}
 \definecolor{WHITE}{gray}{1}
 \definecolor{RED}{rgb}{1,0,0}
 \definecolor{GREEN}{rgb}{0,1,0}
 \definecolor{BLUE}{rgb}{0,0,1}
 \definecolor{CYAN}{cmyk}{1,0,0,0}
 \definecolor{MAGENTA}{cmyk}{0,1,0,0}
 \definecolor{YELLOW}{cmyk}{0,0,1,0}
}

\makeatother

\usepackage{babel}
\begin{document}

\title{Feasibility of giant fiber-optic gyroscopes}

\author{Stephan Schiller}

\affiliation{Institut f\"ur Experimentalphysik, Heinrich-Heine-Universit\"ut D\"usseldorf,
D\"usseldorf, Germany}
\begin{abstract}
The availability of long-distance, underground fiber-optic links opens
a perspective of implementing interferometric fiber-optic gyroscopes
embracing very large areas. We discuss the potential sensitivity,
some disturbances and approaches to overcome them. 
\end{abstract}
\maketitle

\section{Introduction}

Optical gyroscopes are precision instruments widely applied in inertial
guidance of aiplanes, rockets and vessels. The scientific applications
of gyroscopes include geophysical and General Relativity measurements
\cite{Stedman,Bosi,Chow}. Research-type gyroscopes for the measurement
of the fluctuations of the Earth rotation rate $\Omega_{E}$ have
been successfully implemented as active ring laser gyroscopes. In
these, counter-propagating laser waves oscillate and generate a beat
signal. The ring laser gyroscope at Fundamentalstation Wettzell (Germany)
currently exhibits the best long-term stability. It has a ring area
$A\,=4$ ~m$^{2}$ and reaches approx. $1\times10^{-8}$ relative
resolution in $\Omega_{E}$ for integration times of 1 hour to 1 day
\cite{Schreiber,Wettzell web page}. This is made possible by operating
the gyroscope in a (passively) very stable laboratory environment,
and employing active stabilization techniques, which suppress fluctuations
of ring area and of other properties of the gyroscope. 

Since the potential sensitivity of gyroscopes scales with the enclosed
area, early on large gyroscopes were conceived and implemented, in
particular one reported by Michelson and coworkers in 1925 with enclosed
area $A\simeq0.2$~km$^{2}$ and recently a 834~m$^{2}$ ring laser
gyroscope in New Zealand \cite{Hurst}. In interferometric (passive)
fiber-optic gyroscopes \cite{Lefevre}, a large sensitivity may be
achieved in a compact footprint by coiling a fiber with thousands
of turns, yielding areas of tens of m$^{2}$.

It is of interest to consider alternative approaches to the established
ones, for example in view of increasing the number of operating gyroscopes,
so as to improve the combined signal-to-noise ratio of the measurement
of fluctuations of $\Omega_{E}$. The recent development of long-distance,
fiber-based optical frequency transfer links \cite{Terra,Kefelian,Lopez}
provides important information for such approaches. These links use
commercial or research network underground telecommunication fibers
in the 1.5~\textmu{}m spectral range. In the immediate future, it
is intended to use such links to compare atomic clocks located in
different laboratories, in order to characterize their performance
and to set up a network of clocks implementing a future new definition
of the unit of time. It is also foreseen to use comparisons between
distant clocks for local measurements of the gravitational potential
of the Earth, making use of the general-relativistic effect of time
dilation. 

In view of these developments, we discuss here whether existing fiber
links could enable a complementary application, namely gyroscopes.
That is, we consider applying the concept of the small-scale interferometric
fiber gyroscope to an already installed underground fiber network.

\section{Basic Concept}

\subsubsection{Principle}

\begin{figure}
\centering{}\label{Figure: fiber network}\includegraphics[bb=100bp 50bp 630bp 480bp,clip,scale=0.6]{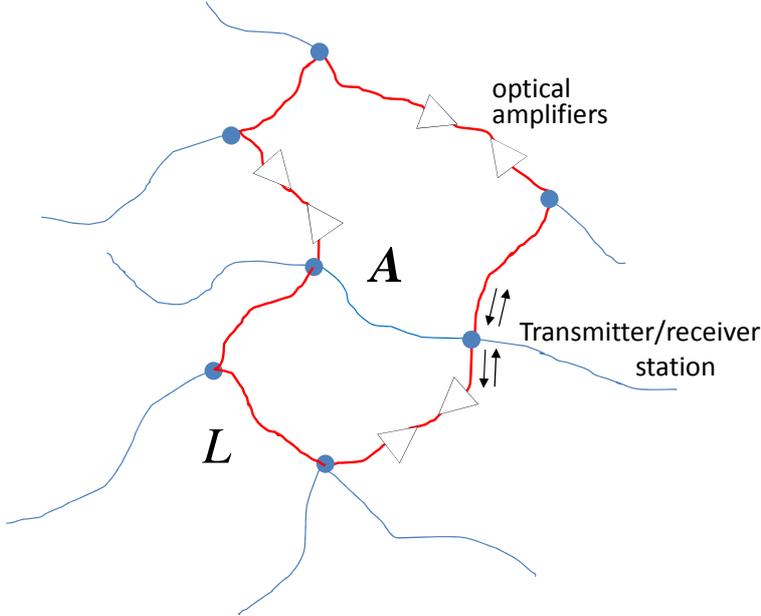}\caption{\label{fig:Schematic-of-a fiber g}(Color online) Schematic of a fiber
gyroscope using part of an existing fiber network (red). A wave produced
by a laser source in the transmitter node is split and sent around
the loop in both directions (black arrows). At the same node the waves
are received and interfered, producing the Sagnac signal. $L,\, A$
are the loop length and enclosed area, respectively.}
\end{figure}

Consider the geometry shown in Figure~\ref{fig:Schematic-of-a fiber g},
which is similar to a usual passive fiber gyroscope. The fiber loop
(ring) would in practice be formed by sections of existing underground
fiber that are selected to form a closed path, enclosing an area $A$.
From a station a (frequency-stable) laser wave of angular frequency
$\omega$ is sent around a fiber loop in the two opposite directions.
If the fiber loop is very long, at intervals of ca.~200 km amplifiers
are required that coherently amplify the two counter-propagating waves.
After their respective round-trips, the two waves are brought to interference.
The interference signal contains the Sagnac phase, 

\[
\Phi_{S}=4\,\omega\,{\bf {A\cdot\Omega}}/c^{2}\ .
\]
which does not depend on the index of refraction of the fiber. For
example, for a circular loop of circumference $L=500$~km, assumed
for simplicity to be perpendicular to ${\bf \Omega}$, and for Earth's
rotation rate $\Omega=\Omega_{E}$~, the Sagnac phase for a wavelength
1.5~\textmu{}m is $\Phi_{E}\simeq8\times10^{5}$~rad. Usually, the
phase itself is not of interest, but its time variations.

One condition for proper operation of such a large-area fiber interferometer
is that the amplifiers introduce only a small nonreciprocal phase
shift. This appears to be fulfilled by using the recently demonstrated
Brillouin amplifiers \cite{Terra}. A second condition is that the
``fast'' fiber length noise, i.e. that due the spectral components
with Fourier frequency larger than $1/T$, where $T$ is the light
propagation time around the fiber, and which would prevent the two
counter-propagating waves from experiencing the same optical path,
is sufficiently averaged out. By using a long enough integration time
$\tau$, the fiber noise will be common to both light waves. This
is discussed in more detail in Section III. B. 

\begin{figure}
\centering{}\includegraphics[scale=0.55]{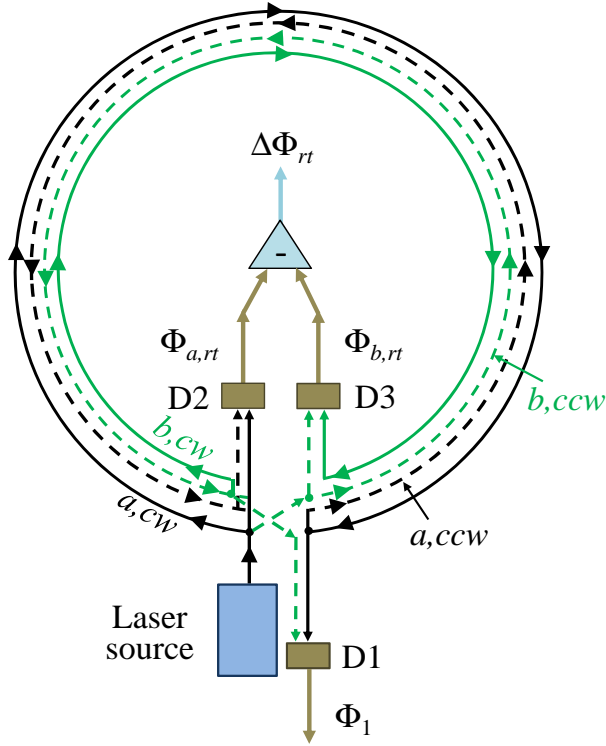}\caption{\label{fig:Optical-layout-of}(Color online) Optical layout of the
fiber gyroscope. Four waves of equal frequency are propagating around
the loop. Full lines denote clockwise propagating waves, dashed lines
counter-clockwise propagating ones. D1, D2, D3 are photodetectors
at which interference of two waves occurs. Their signals contain information
about relative phases and their time variations. $\Phi_{1}$ is the
phase from which the Sagnac phase $\Phi_{S}$ is determined.}
\end{figure}

\subsubsection{Signals}

Figure \ref{fig:Optical-layout-of} shows in more detail the set-up
considered here. Originating from a common laser source, one wave
($a,cw$) is injected clockwise into the loop and one wave ($b,ccw$)
is injected counterclockwise. The phases at the ends of their loops
are

\begin{equation}
\Phi_{a,cw}=\frac{1}{2}\Phi_{S}+\Phi(\omega)+\Phi_{K,a,cw}+\Phi_{T,a,cw}\,,\label{eq:a,cw}
\end{equation}

\begin{equation}
\Phi_{b,ccw}=-\frac{1}{2}\Phi_{S}+\Phi(\omega)+\Phi_{K,b,ccw}+\Phi_{T,b,ccw}\,.\label{eq:b,ccw}
\end{equation}
Here, $\Phi_{S}$ is the Sagnac phase, $\Phi(\omega)=\int_{0}^{L}\beta(\omega)dl$
is the usual propagation phase shift ($\beta$ is the propagation
constant), $\Phi_{K}$ is the Kerr phase shift caused by the presence
of the optical waves having finite intensity, and $\Phi_{T}$ is a
non-reciprocal phase shift of thermal origin (Shupe effect \cite{Shupe}).
At the end of their loop propagations, waves $a,\, b$ are partially
extracted and interfere on a photo-detector D1, producing the Sagnac
signal. It is given by

\begin{equation}
\Phi_{1}=\Phi_{a,cw}-\Phi_{b,ccw}=\Phi_{S}+(\Phi_{K,a,cw}-\Phi_{K,b,ccw})+(\Phi_{T,a,cw}-\Phi_{T,b,ccw})\,.\label{eq:Sagnac signal}
\end{equation}

Note that it is modified by Kerr and Shupe contributions. 

In addition, at the end the loop, each wave is partially reflected
(possibly frequency-shifted by an acousto-optic modulator) and sent
back to the loop beginning. These waves, denoted by $a,ccw$ and $b,cw$~,
experience the phase shifts

\begin{equation}
\Phi_{a,ccw}=-\frac{1}{2}\Phi_{S}+\Phi(\omega)+\Phi_{K,a,ccw}+\Phi_{T,a,ccw}\,,\label{eq:a,ccw}
\end{equation}

\begin{equation}
\Phi_{b,cw}=+\frac{1}{2}\Phi_{S}+\Phi(\omega)+\Phi_{K,b,cw}+\Phi_{T,b,cw}\,,\label{eq:b,cw}
\end{equation}
From these waves we can form two more signals $\Phi_{a,rt},\,\Phi_{b,rt}$
at the detectors D2 and D3, respectively, which carry information
about the round-trip phases,

\begin{equation}
\Phi_{a,rt}=\Phi_{a,cw}+\Phi_{a,ccw}=2\Phi(\omega)+\Phi_{K,a,cw}+\Phi_{K,a,ccw}+\Phi_{T,a,cw}+\Phi_{T,a,ccw}\,,\label{eq:Phi_a,rt}
\end{equation}
\begin{equation}
\Phi_{b,rt}=\Phi_{b,cw}+\Phi_{b,ccw}=2\Phi(\omega)+\Phi_{K,b,cw}+\Phi_{K,b,ccw}+\Phi_{T,b,cw}+\Phi_{T,b,ccw}\,.\label{eq:Phi_b,rt}
\end{equation}
A stabilized fiber is one in which one of these two, say, $\Phi_{a,rt}$
is measured and kept constant in time by active feedback. 

Finally we can form (electronically) the difference round-trip phase

\begin{eqnarray}
\Delta\Phi_{rt} & = & \Phi_{a,rt}-\Phi_{b,rt}\nonumber \\
 & = & (\Phi_{T,a,cw}+\Phi_{T,a,ccw})-(\Phi_{T,b,ccw}+\Phi_{T,b,cw})+(\Phi_{K,a,cw}+\Phi_{K,a,ccw})-(\Phi_{K,b,cw}+\Phi_{K,b,ccw})\,.\label{eq:Delta phi_rt}
\end{eqnarray}
This signal carries information only about disturbances.

\section{Sensitivity}

Consider the state-of-the-art long-buried-fiber link ($L=480\,$km)
of Terra \textit{et al.} \cite{Terra}, at the wavelength 1.5~\textmu{}m.
In effect, they have implemented a device that can measure the Sagnac
phase, although their loop had essentially zero enclosed area. It
is not a conventional Sagnac interferometer with counter-propagating
waves for which the propagation phases cancel, but a single-wave geometry
in which the round-trip propagation phase $\Phi_{a,rt}$ is kept constant
by fiber stabilization. The crucial measurement performed by them
in this context is the phase noise present on the wave traversing
the stabilized link once (i.e. half a round-trip), i.e. on the signal
$\Phi_{a,cw}$. This is equivalent, in absence of the Shupe and Kerr
effect contributions, to one-half the phase noise of the Sagnac signal
$\Phi_{S}$. Terra et al. measured a spectral noise density $S_{D}(f)\simeq2\times10^{-2}\,$rad$^{2}/$Hz
for $0.1\,$Hz $<f<1$~Hz. It may be assumed that this value holds
also for lower frequencies $f$, since the fiber is stabilized. Thus,
the phase error incurred over an integration time $\tau$ is $\sigma_{\Phi}(\tau)\simeq0.14\,{\rm rad}/(\tau/1\,{\rm s})^{1/2}$.
The corresponding relative sensitivity of the Earth rotation rate
measurement is $\sigma_{\Phi}(\tau)/\Phi_{E}\simeq1.8\times10^{-7}/(\tau/1\,{\rm s})^{1/2}$.
If this noise would average down ideally, for a one-day-long integration
($\tau=24$~h) the extrapolated relative resolution of the Earth
rotation rate would be $6\times10^{-10}$. The non-orthogonality of
the loop and Earth rotation axes and the above factor one-half will
moderately increase this number.

It has been predicted \cite{Williams} and confirmed for $L<500\,$km
\cite{Terra} that $S_{D}$ and therefore $\sigma_{\Phi}$ scale as
$L^{3/2}.$ The Sagnac phase, however, scales as $L^{2}$ for a circular
loop. Thus, the relative resolution could improve for longer fiber
links as $L^{-1/2}$, if no unexpected other effects show up.

\section{Disturbances}

\subsection{Stability of the loop area}

The expression for the Sagnac phase shows that for reaching a desired
sensitivity for determining changes in $\Omega$ requires the instability
of the frequency $\omega$ and of the enclosed area $A$ to be correspondingly
small over the desired integration time. The frequency $\omega$ can
be kept constant to the $10^{-15}$ relative level or better by stabilizing
the laser frequency to an atomic clock reference, so this influence
is negligible. However, an active stabilization of the area $A$ does
not seem possible, due to lack of a way of directly determining the
area (independently from measuring the Sagnac effect). The following
considerations indicate the relevance of variations of $A$ and approaches
which may be useful in reducing them. 

The first approach relies on the passive stability of the fiber. Consider
the fiber length $L$ (ring perimeter) stability as an indicator of
enclosed area stability. Fibers that are installed underground are
``quiet'' with respect to their optical length noise. Several groups
have characterized the level of frequency instability achievable in
optical carrier wave transmission along long, unstabilized fiber links.
Terra et al. \cite{Terra} measured the relative frequency instability
$\sigma_{y}(\tau)$ imposed on an optical wave after passage through
a 480~km unstabilized, dark fiber link to be $6\times10^{-15}$ or
less for integration times $\tau$ between 10~s and 500~s, the longest
integration time reported in their study. Kefelian et al. \cite{Kefelian}
observed a roughly constant level below $1\times10^{-14}$ for $\tau$
up to 100 000~s in a 108~km long fiber link. Lopez et al. \cite{Lopez}
measured a similar level for 150~km of telecommuncation fiber carrying
internet data. The corresponding optical phase instability is approximately
$\delta\Phi\simeq\omega\tau\,\sigma(\tau)$. For the first example,
$\delta\Phi\simeq4\times10^{3}$~rad at $\tau=500$~s and correspondingly
higher at longer integration times. Referred to a $L=\,$500 km length,
this implies a relative instability of the optical path length of
$\delta(nL)/nL\simeq\delta\Phi/\Phi\simeq4\times10^{3}/(2\pi\times1.45\times5\times10^{5}\,{\rm m}/(1.5\,\mu{\rm m}))\simeq1\times10^{-9}$
at $\tau=$~500~s ($n$ is the refractive index of the fiber). It
is dominated by fluctuations of the index of refraction of the fiber,
whose temperature sensitivity is significantly larger than that of
the length. At first sight, the above value would indicate a relative
area instability of less than $1\times10^{-9}$ for integration times
up to 500~s. However, there may be area variations not related to
perimeter length variations, e.g. by ground movement. Assume for simplicity
a circular ring. Requiring an area instability $\delta A/A<1\times10^{-9}$
(a necessary condition if a Earth rotation rate sensitivity better
than the best gyroscopes is aimed for) for a ring of $R=100$~km,
leads to the requirement $\delta R<50$~\textmu{}m for the (ring-averaged)
radius change. Such a level may be realistic over sufficiently short
time intervals (minutes to hours). The level of instability of the
area over medium and long averaging times is of fundamental importance
and should be determined experimentally. 

The second approach is an active one. The optical phase accumulated
over the loop (ignoring the Sagnac phase) may be written as

\[
\Phi(\omega)=c^{-1}\omega\int_{0}^{L}n(\omega,s(l),T(l))\, dl\,,
\]
where the dependence of the fiber index on the strain $s$ and the
temperature $T$ along the fiber has been introduced. A Taylor expansion
of the integral for small, but position-dependent variations in the
fiber length, the strain, and the temperature around their respective
mean values $L_{0},\, s_{0},\, T_{0}$ yields the phase deviation
as 

\begin{equation}
\delta\Phi(\omega)\simeq c^{-1}\omega\left(n(\omega,s_{0},T_{0})\delta L+n_{s}(\omega,s_{0},T_{0})\int_{0}^{L}\delta s(l)dl+n_{T}(\omega,s_{0},T_{0})\int_{0}^{L}\delta T(l)dl\right)\,.
\end{equation}
$\delta L$ denotes the overall length variation. The subscripts $s$
and $T$ indicate partial derivatives of the refractive index with
respect to strain and temperature, respectively. Note that the refractive
index depends on the optical frequency $\omega$, but its derivatives
do so as well. Consider sending three sufficiently spaced frequencies
$\omega_{1},\,\omega_{2},\,\omega_{3}$ (which should lie in the spectral
region in which the fiber has low propagation loss and for which optical
amplifiers are available) around the ring and measuring the three
corresponding phase deviations $\delta\Phi(\omega_{1}),\,\delta\Phi(\omega_{2}),\,\delta\Phi(\omega_{3})$
as a function of time. In order not to accumulate Sagnac phases, the
measurement is done by the sending the waves around the ring and back
the same way (as in fiber link stabilization procedures), whereby
each total phase is twice the single-turn phase (after fast fluctuations
are averaged out). The corresponding signal is given in Eqs.(\ref{eq:Phi_a,rt})
or (\ref{eq:Phi_b,rt}). By linear combination of the three phase
deviations $\delta\Phi(\omega_{i})$ with weights determined by the
refractive index partial derivatives, we may extract information about
the loop length variation $\delta L$. This variation may be actively
compensated using e.g. a special piece of fiber being part of the
ring, whose temperature or strain can be actively modified. Based
on the achievements in long-distance fiber links it is expected that
the fiber geometrical length $L$ can be stabilized to a relative
level far below $1\times10^{-12}$ for all relevant integration times.
This stabilization could be helpful in reducing also the area fluctuations
and this hypothesis should be tested experimentally. However, the
link stabilization cannot compensate variations of the encompassed
area $A$ that are not due to variations of the link length $L$.
It should be noted that this approach is conceptually related to the
active perimeter stabilization implemented in the large ring laser
gyroscope \cite{Schreiber}. 

\begin{figure}
\centering{}\label{Figure: more complicated geometries}\includegraphics[bb=0bp 0bp 567bp 567bp,clip,scale=0.45]{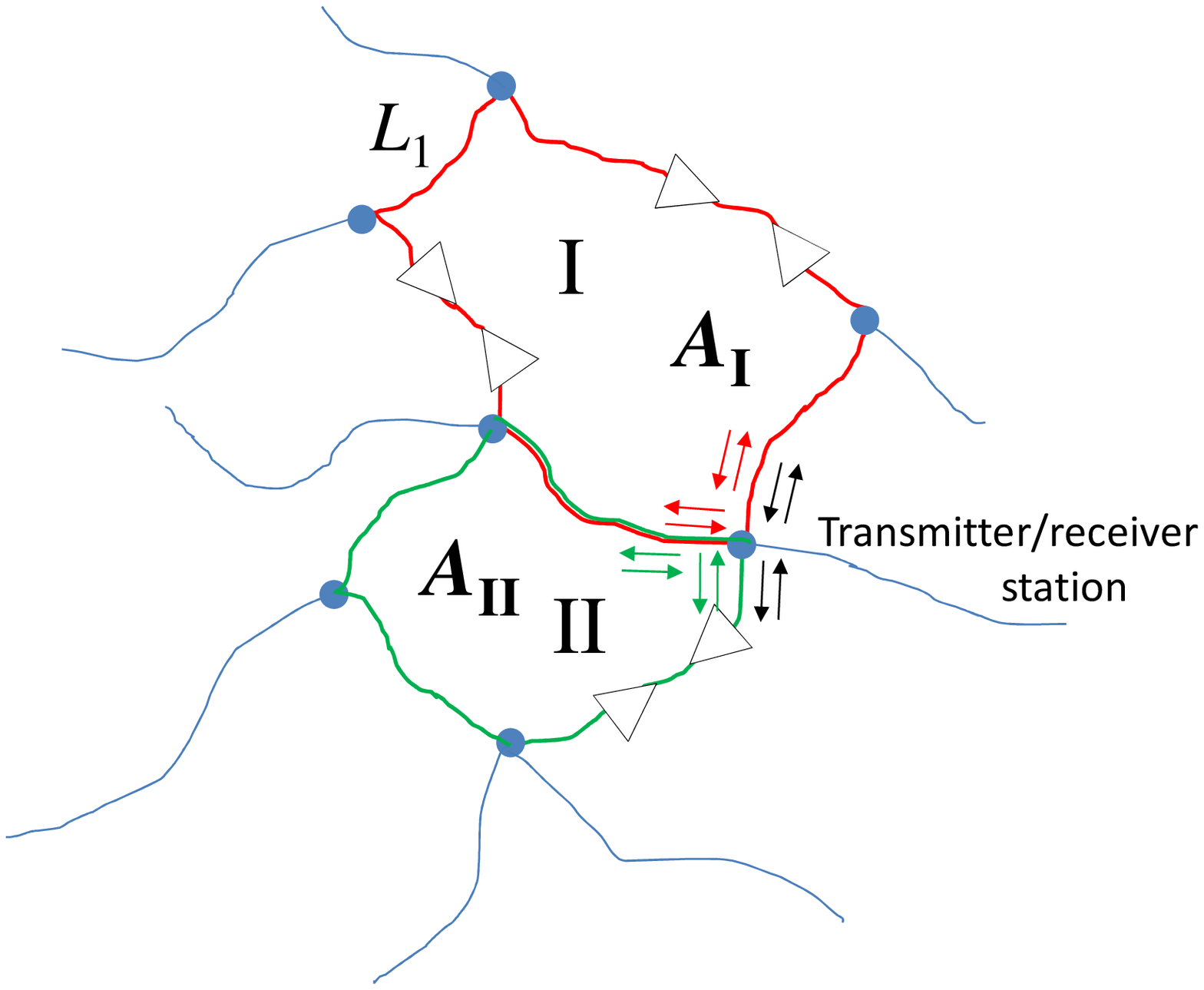}\includegraphics[bb=0bp 0bp 567bp 567bp,clip,scale=0.45]{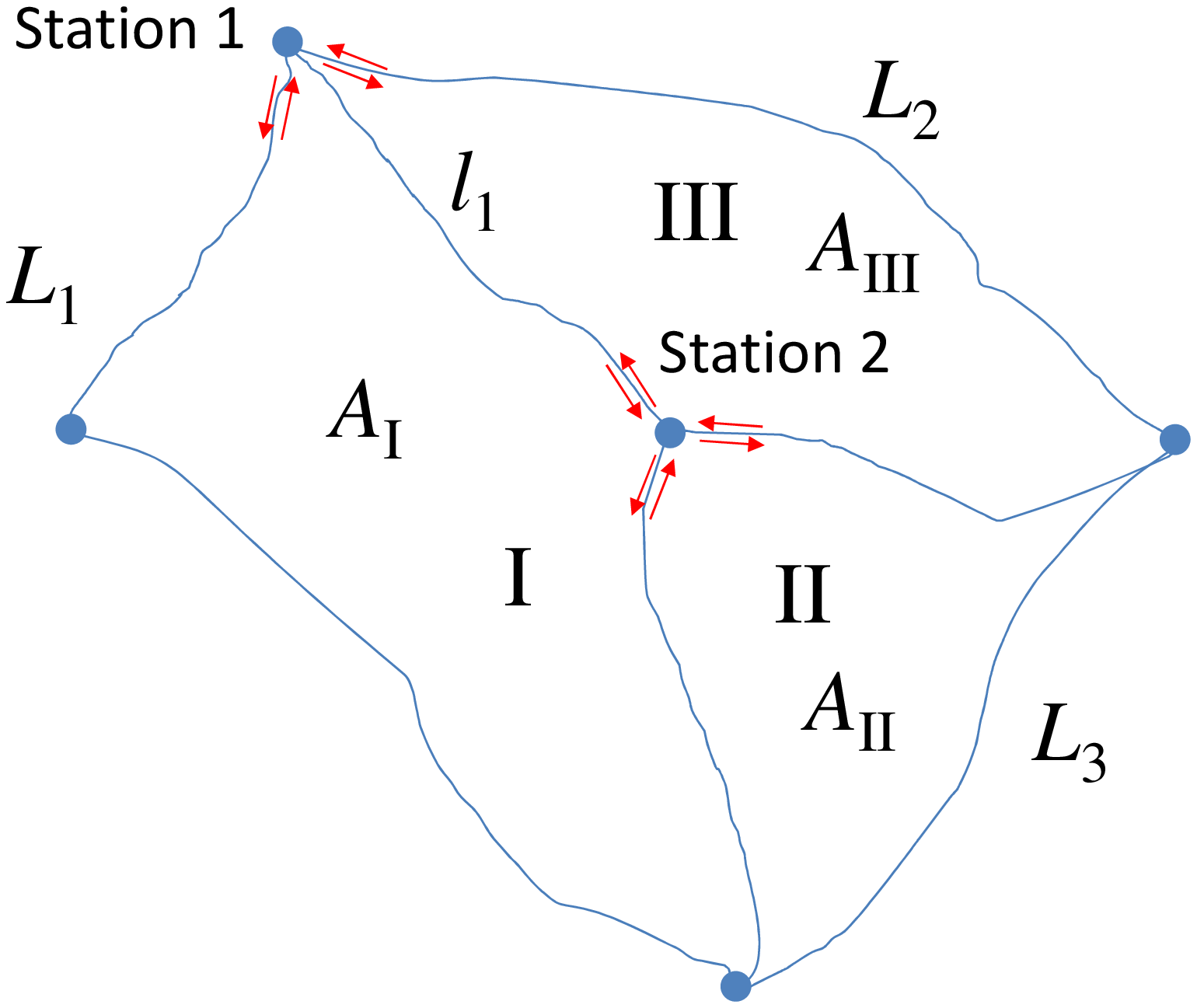}\caption{\label{fig:Proposed-fiber-network}(Color online) Proposed fiber network
geometries for the discrimination of area changes. Left: simplest
geometry with three gyroscopes I, II, and I-II. The laser waves traveling
in these three gyroscopes from the single station are indicated by
the red, green, and black arrows, respectively. Right: more complex
geometry. Station 1 operates the outer fiber loop enclosing areas
$A_{I},\, A_{II},\, A_{III}$. Station 2 operates (possibly alternating)
the three smaller individual loops I, II, III. Laser waves are indicated
by arrows, but are not color-coded. For example, a change in the outer
loop area by deformation of the outer loop link subsection $L_{1}$
can be detected and rejected by correlation with of the Sagnac signal
from gyroscope I and non-correlation with signals from gyroscopes
II and III. If instead a deformation of the link subsection $l_{1}$
occurs this would not appear in the outer gyroscope signal, but can
nevertheless be detected by anti-correlation between the signals from
gyroscopes I and III and noncorrelation between the signals from gyroscope
I and gyroscope II.}
\end{figure}

We suggest as a third option a more complicated link network to reject
such area variations, shown in Figure~\ref{fig:Proposed-fiber-network}
(left). In it, one link is added which splits the original area $A$
in two approximately equal parts $A_{I},\, A_{II}$. This results
in three gyroscopes, I, II, and the combined loop I-II, which can
be operated independently, either simultaneously or alternating. Consider
now a change in area of the combined gyroscope I-II due to a spatial
displacement of link section $L_{1}$. This will result in a signal
in the I-II gyroscope, but also in the smaller gyroscope I, while
not in the gyroscope II. By detecting such a signature, the signal
on the I-II gyroscope would be identified as a disturbance. A similar
reasoning applies if the I-II gyroscope area changes due displacement
of other sections. In contrast, a change in rotation rate $\Omega$
yields a common (but unequal) signal on all three gyroscopes. The
sensitivities of the three gyroscopes to rotation differ because of
the different areas, but these areas can be computed from a measurement
of the geometry (e.g. using GPS) and taken into account in the disturbance
rejection analysis. 

More complicated geometries can be considered, e.g. the one shown
in Figure~\ref{fig:Proposed-fiber-network} (right), containing two
stations and four gyroscopes. With such geometries, the rejection
of disturbances might be even more effective.

In addition to the instability of the enclosed area we point out that
the Sagnac phase contains the scalar product ${\bf {A\cdot\Omega}}$.
Variations in the orientation of the fiber loop with respect to the
Earth axis will thus also produce signals. These may be undesired
for some applications but possibly useful for others. 

\subsection{Nonreciprocal errors}

\subsubsection{\label{sub:Shupe-effect}Shupe effect}

The Shupe effect \cite{Shupe} is caused by the presence of a time-dependent
inhomogeneous temperature distribution in the fiber. In a conventional
fiber-optic gyroscope, there are two counter-propagating waves. The
wave sections reaching the detector at the same time have crossed
every section of the fiber loop, except the one at the center, at
a different time. At the two times of passage the respective values
of the temperature will differ (due to temperature drift), causing
a different phase shift for the two waves due to the finite thermo-optic
coefficient $dn/dT$ and the fiber thermal expansion coefficient $\alpha$.
The differential phase accumulates over the length of the fiber, giving
rise to a non-reciprocal phase shift $\Phi_{T}$ which adds to the
Sagnac phase. As the temperature drift $dT/dt$ is not constant in
time, $\Phi_{T}$ and therefore the total phase drifts. According
to the model given by Shupe, the time interval $\tau$ between the
passages of the clockwise and counterclockwise wave through a fiber
section located at a distance $l$ from one end is given by $\tau_{cc-ccw}=\beta(2\ l-L)/\omega$,
where $\beta$ is the propagation constant. Then, the total nonreciprocal
phase is 

\begin{eqnarray}
\Phi_{1,T}=\Phi_{T,a,cw}-\Phi_{T,b,ccw} & \simeq & k(dn/dT+n\alpha)\int_{0}^{L}\tau_{cc-ccw}(dT(l)/dt)dl\nonumber \\
 & \simeq & (n\omega/c^{2})(dn/dT+n\alpha)\int_{0}^{L}(2l-L)(dT(l)/dt)dl\,.\label{eq:Phi_1,T}
\end{eqnarray}
Here, $\beta\simeq k\, n=n\,\omega/c$ has been used. Note that $\Phi_{1,T}$
vanishes if the temperature change rate is constant over the fiber
length. A model assumption for the inhomogeneous temperature change
in the fiber discussed by Shupe is $dT(l)/dt=(\Delta T/\Delta t)(l/L)$~.
Shupe proposed and studies have been performed on particular winding
(coiling) geometries in order to reduce the effect. These special
geometries are obviously not applicable here. 

In order to model the Shupe effect for a long buried fiber, we divide
the length $L$ into $N$ intervals, each of which has a particular
temperature drift rate uncorrelated with that of the neighboring intervals.
The temperature drift rates are assumed to have a Gaussian random
distribution with zero mean and standard deviation $\sigma_{dT/dt}=0.1\,$K/h,
a typical value for buried fibers \cite{Terra}. A numerical modeling
shows that the total phase also has a Gaussian distribution with zero
mean, and a standard deviation approximately given by $\sigma_{T}\simeq\sigma_{dT/dt}\sqrt{N}L^{2}k\,(dn/dT+n\alpha)/c$~.
For a loop length $L=500\,$km and $N=500$ intervals, $\sigma_{T}\simeq$
22~rad. This is a large value and time-varying temperature drift
rates will overwhelm any Earth rotation rate fluctuation signal. 

Consider now the configuration used in fiber noise cancellation (Figure
\ref{fig:Optical-layout-of}): a wave running in one direction around
the loop, reflected back at the end and returning to the emitter.
Now, the sum of time delays for reaching the same fiber section $l$
on the forward and backward trips, is always $\beta(2L)/\omega$~.
Therefore $\Phi_{T,a,cw}+\Phi_{T,a,ccw}=\Phi_{T,b,cw}+\Phi_{T,b,ccw}\propto(\beta(2L)/\omega)\int_{0}^{L}(dT(l)/dt)dl$~.
The difference $\Delta\Phi_{rt}$~between the two round-trip phases,
introduced above in Eq.(\ref{eq:Delta phi_rt}), is therefore independent
of the Shupe effect and only contains information about the Kerr effect. 

For concreteness, we explicitly state the Shupe effect contributions
for the four one-way propagation phases introduced in Eqs.(\ref{eq:a,cw},\ref{eq:b,ccw},\ref{eq:a,ccw},\ref{eq:b,cw}),

\begin{eqnarray}
\Phi_{a,cw}(\omega) & = & \frac{1}{2}\Phi_{S}+\Phi(\omega)+\xi\int_{0}^{L}l\, T_{t}\, dl+\Phi_{K,a,cw}\,,\nonumber \\
\Phi_{b,ccw}(\omega) & = & -\frac{1}{2}\Phi_{S}+\Phi(\omega)+\xi\int_{0}^{L}(L-l)T_{t}\, dl+\Phi_{K,b,ccw}\,.\nonumber \\
\Phi_{a,ccw}(\omega) & = & -\frac{1}{2}\Phi_{S}+\Phi(\omega)+\xi\int_{0}^{L}(2L-l)T_{t}\, dl+\Phi_{K,a,ccw}\,,\nonumber \\
\Phi_{b,cw}(\omega) & = & +\frac{1}{2}\Phi_{S}+\Phi(\omega)+\xi\int_{0}^{L}(L+l)T_{t}\, dl+\Phi_{K,b,cw}\,.\label{eq:test}
\end{eqnarray}

Here we have introduced the parameter $\xi(\omega)=(n(\omega)\omega/c^{2})(dn(\omega)/dT+n(\omega)\alpha)$,
and the short-hand notation $T_{t}=dT(l)/dt$~. $\Phi_{K}$ are Kerr
phases discussed further below.  We see that $\Phi_{S}/2$ always
occurs in combination with $\xi\int lT_{t}dl$~. This part of the
Shupe contribution is therefore always measured together with the
Sagnac phase. It is not possible to determine $\Phi_{S}$ alone from
any linear combination of the above four phases.

Returning to the Sagnac signal, Eq.(\ref{eq:Sagnac signal}), we ignore
the Kerr effect for the time being and have

\begin{equation}
\Phi_{1}(\omega)=\Phi_{a,cw}(\omega)-\Phi_{b,ccw}(\omega)=\Phi_{S}+\xi(\omega)\int_{0}^{L}(2l-L)T_{t}dl\,.
\end{equation}

Although both $\Phi_{S}$ and $\xi(\omega)$ are proportional to $\omega$,
the latter coefficient has an additional (weak) dependence on $\omega$
through the factor $n(\omega)(dn(\omega)/dT+n(\omega)\alpha)$. Therefore,
we can again consider the possibility of using waves of different
frequencies $\omega$. With two frequencies, we may produce the signals
$\Phi_{1}(\omega_{1})$ and $\Phi_{1}(\omega_{2})$~. They are not
linearly dependent, and we may produce an appropriate linear combination
and extract the Sagnac phase $\Phi_{S}(\omega)$ and the Shupe contribution$\int_{0}^{L}(2l-L)T_{t}\, dl$
independently. 

Thus, it is in principle possible to measure the Shupe effect of the
standard Sagnac configuration by measuring at different optical frequencies,
at least if we neglect the influence of the Kerr effect. It therefore
should be possible to correct for the Shupe effect, although to what
level is a question that experiments have to answer. This correction
is compatible with active stabilization of one of the two round-trip
phases.

\subsubsection{Optical Kerr effect}

The intensity of any wave propagating inside a fiber changes the refractive
index via the optical Kerr effect. In the standard Sagnac geometry
with two counter-propagating waves of powers $P_{cw},$ $P_{ccw},$
there arises a nonzero differential phase shift $\Phi_{K,cw}-\Phi_{K,ccw}$
if the average powers of the two waves are not equal, since for each
fiber length interval $dl$, there arises both a self-Kerr phase shift
and a cross-Kerr phase shift (which has an additional factor 2), given
by $d\Phi_{K,cw}=\kappa(P_{cw}+2\, P_{ccw})dl,$ $d\Phi_{K,ccw}=\kappa(P_{ccw}+2\, P_{cw})dl,$
where $\kappa$ is the Kerr coefficient \cite{Chow}. Obviously, the
solution to this problem consists in equalizing the two powers $P_{cw},\, P_{ccw}$
by appropriate means and this has been implemented experimentally
by several groups \cite{Chow}. Another approach, recently demonstrated
\cite{Dangui}, is the use of a photonic crystal fiber instead of
a standard fiber, but this solution is not compatible with already
installed fibers and also exhibits much higher fiber loss. For a long-distance
fiber link, where optical amplifiers are present, the situation becomes
more complex. For example, in the approach of using Brillouin amplification
\cite{Terra}, the power $P^{(p)}$ of the pump waves injected into
the fiber (of order tens of mW) also induces a change of the index
of refraction. However, the effect on the phases of the two counter-propagating
waves is equal because it is a cross-Kerr effect, $\Phi_{K,cw}^{(p)}=\Phi_{K,ccw}^{(p)}=\Phi_{K}^{(p)}=\kappa\int_{0}^{L}2P^{(p)}(l)dl$
and this particular Kerr contribution cancels in the standard Sagnac
geometry.  

We now extend the discussion to the configuration of Fig.~\ref{fig:Optical-layout-of}.
We take into account that in each fiber section there are five waves
present: the pump wave ($p$) (there is no need to differentiate the
contributions in cw and in ccw direction, only the total power is
considered) and the four ``signal'' waves $a,cw\,,\, a,ccw\,,\, b,cw\,,$~and
$b,ccw\,$, each with arbitrary position-dependent powers. Therefore,

\[
\Phi_{K,a,cw}=\kappa\int_{0}^{L}(P_{a,cw}+2P_{a,ccw}+2P_{b,cw}+2P_{b,ccw}+2P^{(p)})dl\,,
\]
and analogously for the other three signal waves. The Kerr contributions
to the Sagnac signals are therefore

\begin{equation}
\Phi_{1,K}=\Phi_{K,a,cw}-\Phi_{K,b,ccw}=\kappa\int_{0}^{L}(-P_{a,cw}+P_{b,ccw})dl\,.
\end{equation}

\begin{equation}
\Phi_{2,K}=\Phi_{K,b,cw}-\Phi_{K,a,ccw}=\kappa\int_{0}^{L}(-P_{b,cw}+P_{a,ccw})dl\,.
\end{equation}
For the round-trip phase difference introduced in Eq.(\ref{eq:Delta phi_rt}),
we have
\begin{eqnarray}
\Delta\Phi_{rt} & = & (\Phi_{K,a,cw}+\Phi_{K,a,ccw})-(\Phi_{K,b,cw}+\Phi_{K,b,ccw})\nonumber \\
 & = & \kappa\int_{0}^{L}(-P_{a,cw}-P_{a,ccw}+P_{b,cw}+P_{b,ccw})dl\,.\label{eq:Delta Phi_rt explicit power dependence}
\end{eqnarray}

This signal is certainly useful for monitoring how large the fluctuations
of the average powers are. 

How to perform active stabilization of one of the Kerr contributions,
say $\Phi_{1,K}$? Ideally, we would measure and control both length-averaged
powers $\int_{0}^{L}P_{a,cw}dl$ and $\int_{0}^{L}P_{b,ccw}dl$. 
But these average values appear not to be easily measurable. For example,
if we turn off alternating either wave $a,cw$ or $b,ccw$ in order
to determine its effect in the Sagnac signal $\Phi_{1}$ or on $\Delta\Phi_{rt}$~,
we also loose these signals. We may instead reduce the input power
of the $a,cw$ and of the $b,ccw$ waves by a fixed fraction (as set
by an accurate power measurement). This will not necessarily reduce
the average power by the same fraction, since there may be saturation
effects inside the fiber due to the amplification stages. Nevertheless,
the difference in the signal with full and reduced power might be
used as the quantity that is kept constant by acting back on the input
power. 

Another option is to stabilize the $a,cw$ and $b,ccw$ powers exiting
the fibers and reaching the detectors. A test of the suitability of
this approach could be performed by implementing this stabilization
for the four waves appearing in Eq.(\ref{eq:Delta Phi_rt explicit power dependence}),
and characterizing the resulting stability of $\Delta\Phi_{rt}$. 

It seems possible to extend these procedures to the case of use of
multiple frequencies considered in Sec. \ref{sub:Shupe-effect}.

In another potential solution, one is led to measure the powers at
several locations along the loop and stabilize their values actively,
e.g. by using a feedback servo on the amplifier pump source closest
to the measurement point (note that Brillouin amplification is unidirectional)
or to the laser source. While it is possible to distinguish between
the $cw$ and $ccw$ waves using directional couplers, it is less
straightforward to distinguish between the waves $a$ and $b$ running
in the same direction, as they have nearly the same frequency. One
possibility consists in amplitude-modulating with different frequencies
the two waves $a,cw$ and $b,ccw$ before entering the fiber, and
detecting the corresponding power modulation signals.

\subsubsection{Other effects}

Other noise sources exist in fiber gyroscopes, such as Rayleigh back-scattering.
The scaling derived for usual fiber gyroscopes \cite{Chow} indicates
that the effect is strongly reduced for large rings. The link stabilization
technique will compensate for changes in the effect arising from the
round-trip wave propagation, but the rotation signal will have a contribution
that depends only on the counterclockwise wave propagation. Thus the
stabilization technique does not appear to provide a means to compensate
for the effect. However, modulation techniques that have been suggested
for fiber gyroscopes should also be applicable to giant rings.

\section*{Conclusion}

In this note, we proposed the use of optical fibers belonging to installed
underground fiber networks for an implementation of gyroscopes with
potentially high sensitivity, thanks to the giant encompassed area,
on the order of $10^{10}$~m$^{2}$. The stability of the gyroscope
is of crucial importance if a high sensitivity is to be attained.
While a study of the perturbations affecting large ring laser structures
has uncovered difficulties \cite{Hurst}, it is here suggested that
in a stabilized, buried fiber-based gyroscope the perturbations may
be of different type. The most basic requirements for the implementation
were discussed. 

The need for optical amplifiers that do not produce significant non-reciprocal
time-varying phases was pointed out but seems feasible. 

It was furthermore pointed out that the Shupe effect may be measured
using a multi-frequency approach, allowing to correct the Sagnac signal
which is affected by it. No elegant measurement concept for determining
the Kerr effect was found, so that an indirect technique is required
instead. Approaches considered for achieving an approximate stabilization
of the relevant loop-averaged powers are a point-wise stabilization
of each wave's power, or an output power stabilization, or a stabilization
of the round-trip phase difference. 

A further crucial requirement is a time-stable enclosed area $A$.
The level of instability of large fiber loop areas has not been measured
so far, to the author's knowledge, and and should be determined experimentally.
A gyroscope that aims at being a competitive instrument for Earth
rotation rate studies requires a relative area instability of $1\times10^{-9}$
or less (i.e. an area variation not larger than on the order of 10~m$^{2}$
for a giant fiber loop). It is an open issue whether this is achievable,
but even a much higher instability could lead to useful applications,
since a giant fiber gyroscope implements an averaging over a larger
area, and thus a different kind of measurement as compared to conventional
instruments. In the best case, the passive stability of buried fibers
will provide sufficient area stability due to averaging of disturbances
over a large spatial scale, in particular if the measurement can be
performed within a sufficiently short time. This would come at the
expense of sensitivity, however, since a short averaging time does
not allow to average down the noise significantly. 

In another, still favorable case, there might be a partial correlation
between area change and geometric path length change, in which case
a measurement of length fluctuations and their active compensation
or correction could represent a partial solution. Extensions of long-distance
fiber link stabilization techniques, aiming to stabilize the geometric
path length $L$ instead of the optical path length $n\, L$ are suggested
for this purpose. 

As a third option, more complex fiber network gyroscope geometries
might be helpful for spurious signal rejection due to area changes.
They require more, but not fundamentally different infrastructure
as compared to the simplest ring geometry. The latter two approaches
may be helpful in achieving low gyroscope instability also on long
time scales. 

Finally, it is suggested that experimental studies are undertaken
to verify the concepts proposed here. To be specific, it is desirable
to extend the noise measurements performed by Terra et al. to lower
frequencies; to verify the suggestion of dual frequency use for extraction
of the Shupe effect and the use of three frequencies for length stabilization;
to measure the Kerr-effect-induced phase fluctuations, and to verify
the feasibility of the power stabilization procedures sketched here.
\medskip{}

\textit{Note:} During the finalization of the revised version of this
manuscript, the first experimental realization of a large-area fiber
Sagnac interferometer using existing telecom fiber links has been
reported by Clivati \textit{et al}, arxiv:1212.5717. The 20~km\textsuperscript{2}
area gyroscope exhibits a noise floor of $\sigma_{\Phi_{S}}(\tau>20\,\hbox{\rm s})/\Phi_{E}\simeq4\times10^{-5}$
relative to the Earth rotation rate.
\begin{acknowledgments}
I thank A.Yu. Nevsky and I. Ernsting for comments on the manuscript. \end{acknowledgments}

\end{document}